\documentclass[10pt,english]{article}
\usepackage{mathptmx}
\usepackage[T1]{fontenc}
\usepackage[latin9]{inputenc}
\usepackage{geometry}
\usepackage{color}
\usepackage{babel}
\usepackage{float}
\usepackage{amsmath}
\usepackage{amsthm}
\usepackage{graphicx}
\usepackage{esint}
\usepackage[authoryear]{natbib}
\usepackage[unicode=true,pdfusetitle,
 bookmarks=true,bookmarksnumbered=true,bookmarksopen=true,bookmarksopenlevel=13,
 breaklinks=false,pdfborder={0 0 0},pdfborderstyle={},backref=false,colorlinks=true]
 {hyperref}

\makeatletter
\newcommand{\lyxaddress}[1]{
	\par {\raggedright #1
	\vspace{1.4em}
	\noindent\par}
}

\@ifundefined{date}{}{\date{}}
\usepackage{color}
\usepackage[usenames,dvipsnames,svgnames,table]{xcolor}
\usepackage{multicol}

\definecolor{burgundy}{rgb}{0.5, 0.0, 0.13}
\definecolor{airforceblue}{rgb}{0.36, 0.54, 0.66}
\hypersetup{urlcolor=airforceblue}
\hypersetup{citecolor=burgundy}

\setcitestyle{notesep={, },super,open={},close={},comma,aysep={,},yysep={,}}  
 \makeatletter
   \renewcommand\@biblabel[1]{#1.}
 \makeatother

\makeatother

\begin{document}
\title{The number and location of Jupiter's circumpolar cyclones explained
by vorticity dynamics}
\author{\href{https://orcid.org/0000-0002-3645-0383}{Nimrod Gavriel}$^{1\star}$
and \href{https://orcid.org/0000-0003-4089-0020}{Yohai Kaspi}$^{1}$}
\maketitle

\lyxaddress{\begin{center}
\textit{$^{1}$Department of Earth and Planetary Sciences, Weizmann
Institute of Science, Rehovot, Israel}\\
\textit{$^{\star}$\href{mailto:nimrod.gavriel@weizmann.ac.il}{nimrod.gavriel@weizmann.ac.il}}
\par\end{center}}

\lyxaddress{\begin{center}
Preprint May 27, 2021\textit{}\\
\textit{Nat. Geosci}. 14, 559--563 (2021). DOI:\href{https://doi.org/10.1038/s41561-021-00781-6}{10.1038/s41561-021-00781-6}
\\
(Received September 7, 2020; Revised February 25, 2021; Accepted May
28, 2021) 
\par\end{center}}
\begin{abstract}
The Juno mission observed that both poles of Jupiter have polar cyclones
that are surrounded by a ring of circumpolar cyclones. The North Pole
holds eight circumpolar cyclones and the South Pole possesses five,
with both circumpolar rings positioned along latitude $\sim84^{\circ}$~N/S.
Here we explain the location, stability, and number of the Jovian
circumpolar cyclones by establishing the primary forces that act on
them, which develop because of vorticity gradients in the background
of a cyclone. In the meridional direction, the background vorticity
varies owning to the planetary sphericity and the presence of the
polar cyclone. In the zonal direction, the vorticity varies by the
presence of adjacent cyclones in the ring. Our analysis successfully
predicts the latitude and number of circumpolar cyclones for both
poles, according to the size and spin of the respective polar cyclone.
Moreover, the analysis successfully predicts that Jupiter can hold
circumpolar cyclones while Saturn currently cannot. The match between
the theory and observations implies that vortices in the polar regions
of the giant planets are largely governed by barotropic dynamics,
and that the movement of other vortices at high-latitudes is also
driven by interaction with the background vorticity.
\end{abstract}
\twocolumn 

\begin{figure}
\centering{}\includegraphics[width=1\columnwidth]{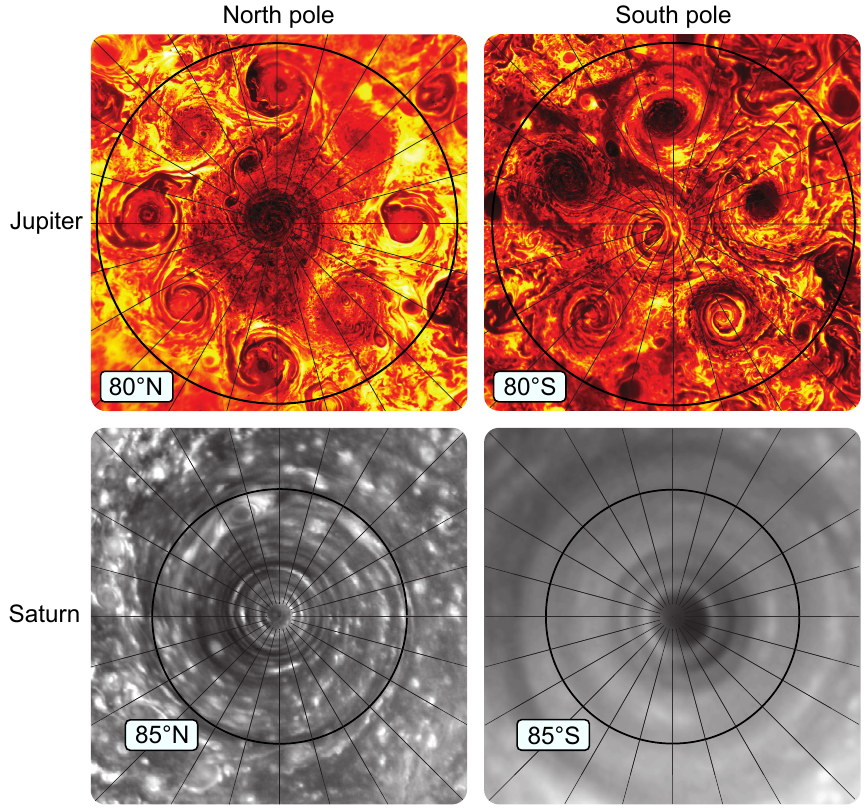}\caption{\textbf{Observations of the polar and circumpolar cyclones of Jupiter
and Saturn.} The images of Jupiter are infrared measurements taken
by Juno's JIRAM camera\citep{adriani2018} (adapted with permission).
The images of Saturn were captured by the Cassini ISS \citep{sayanagi2017cassini}
(adapted with permission). Longitude lines are pointing to the poles
and are $15^{\circ}$ apart. For Jupiter, longitude $0^{\circ}$ in
System III is positioned at the center right of the images. A latitude
circle is shown at $80^{\circ}$~N/S for Jupiter and at $85^{\circ}$~N/S
for Saturn. \label{fig: Infrared-images-of jupiter poles}}
\end{figure}

The Juno orbiter entered in 2016 into a polar 53-day orbit around
Jupiter\citep{bolton2017jupiter} and transmitted the first-ever detailed
observations of its poles\citep{orton2017first,adriani2018} (Fig.~\ref{fig: Infrared-images-of jupiter poles}).
The images showed a unique configuration of cyclonic vortices at both
poles\citep{adriani2018}. Each pole contains a polar cyclone (PC)
with its center positioned close to the pole and is surrounded by
a ring of circumpolar cyclones (CPCs). The ring incorporates eight
cyclones around the north pole and five cyclones around the south
pole\citep{adriani2018}. These constellations of cyclones are very
stable and only drifted slightly with no substantial changes in vortex
morphologies and sizes for over two years\citep{tabataba2020long,adriani2020two}.
The observable diameters of the polar and circumpolar cyclones range
between $4,000$ and $6,000$~km\citep{adriani2020two}, where the
velocities inside the cyclones reach up to $100\,\text{ms}^{-1}$\citep{grassi2018first}.
Such vortex-crystal formations were previously predicted as a 2D relaxation
mechanism of turbulence and were found in experiments that simulate
2D flow with magnetized electron columns\citep{fine1995relaxation,schecter1999vortex}.
However, these configurations were never observed in nature besides
at the poles of Jupiter.

Saturn is generally similar to Jupiter and displays comparable dynamics\citep{showman2018global,kaspi2020comparison,galanti2019}.
Both poles of Saturn also contain a central polar cyclone, each faster
and more extensive than each of the Jovian PCs\citep{baines2009saturn}.
However, neither of the Saturnian poles have an observable circumpolar
ring of vortices. Any theory for the Jovian polar CPCs must also explain
the absence of CPCs on Saturn.

For the poles of Jupiter, the polygonal structure of the CPCs was
modeled using a single-layer shallow water (SW) model\citep{li2020}
to investigate their depth and structure and provide evidence for
anticyclonic shielding around the CPCs, which was shown to be necessary
to inhibit cyclone mergers.\textbf{\textcolor{black}{{} }}\textcolor{black}{In
another approach, it was shown, using deep 3D models, that Jovian
cyclones could extend deep and may emanate from convection of heat\citep{yadav2020deep,yadavbloxam2020deep,garcia2020deep}
and drift poleward}\citep{afanasyev2018cyclonic,afanasyev2020poleward}\textcolor{black}{.
Such 3D vortex behavior was also studied in laboratory experiments\citep{aubert2012,lemasquerier2020remote}.}

A primary mechanism fundamental to vortex dynamics is a secondary
drift that acts on a vortex due to the sphericity of the planet\citep{rossby1948displacements,adem1956series},
and is commonly known as a ``$\beta$-drift''. The direction of
the drift is determined by the sign of the vortex and by the direction
in which the background planetary vorticity rises, resulting in selective
transport of cyclonic anomalies poleward while anticyclonic anomalies
move equatorward. This mechanism is a major contributor to the poleward
migration of tropical cyclones on Earth\citep{smith1990numerical,shapiro1992hurricane,smith1993hurricane,franklin1996tropical,chan2005physics}.
Using a 2.5-layer SW model, it was shown that moist convection, formed
beneath the cloud level of a gas giant, can generate such cyclonic
anomalies\citep{oneill2015polar,oneill2016}. These, in turn, converge
at the poles due to the $\beta$-drift and can lead to coherent PCs
such as those on Saturn\citep{scott2011polar,oneill2015polar,brueshaber2019dynamical}.
In addition to the $\beta$-drift, an equivalent mechanism can drive
a vortex by the presence of any background vorticity gradient. For
example, this gradient can be induced by jet streams and influence
the movement of a crossing vortex\citep{zhao2009observational,riviere2012potential}.
This generalization appears to be crucial for understanding the stability
of the circumpolar vortices on Jupiter.

\subsubsection*{Vortex drift by a background vorticity gradient}

\begin{figure}
\begin{centering}
\includegraphics[width=1\columnwidth]{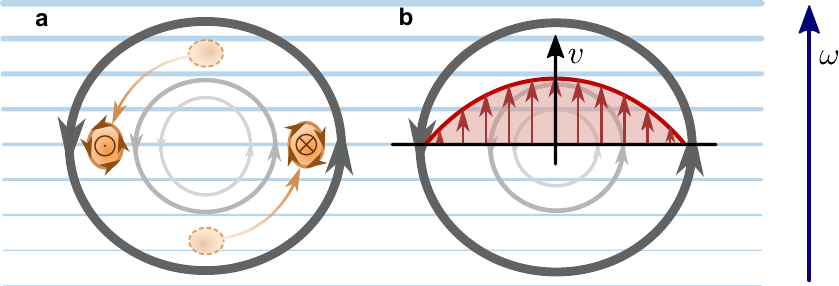}
\par\end{centering}
\caption{\textbf{Beta-drift schematic. }Gray contours represent streamlines
of a vortex. Blue contours represent lines of equal background vorticity
($\omega$), where their increasing thickness represents the gradual
increase in the magnitude of the background vorticity. \textbf{a},
The Lagrangian motion of two fluid parcels leads to a dipole of induced
vorticity due to the conservation of PV during the motion led by the
vortex. \textbf{b}, A velocity profile induced in the vortex by the
shearing due to the vorticity dipole illustrated in \textbf{a}.\label{fig: beta-drift-fig}}
\end{figure}

For understanding the stability of the CPCs, it is essential to generalize
the beta-drift as a force that acts in the direction of rising background
vorticity. For intuition, one can think of a simplistic scenario in
which the conserved potential vorticity (PV) is a superposition of
constant vorticity by a solid disk, counter-clockwise rotating vortex,
and of an unspecified background vorticity ($\omega$ in Fig.\ \ref{fig: beta-drift-fig})
that monotonically rises northward. A fluid parcel that starts on
the southern edge of the vortex, where background vorticity is small,
will be carried by the vortex circulation to the eastern edge, where
background vorticity is higher (Fig.~\ref{fig: beta-drift-fig}a).
To conserve PV, negative relative vorticity would be induced. The
opposite will happen with a northern parcel that will induce positive
vorticity while reaching the west side. This dipole of induced vorticity,
usually termed ``$\beta$-gyres''\citep{fiorino1989some,sutyrin1994intense},
will then shear the velocity field and thus generate a northward velocity
profile (Fig.~\ref{fig: beta-drift-fig}b). The mean force acting
on a vortex due to this phenomenon is proportional to the gradient
of background vorticity both in magnitude and in direction. Counter-clockwise
rotating vortices are pulled toward the highest ascent of background
vorticity, while clockwise rotating vortices are pulled toward the
highest descent. In the planetary context (considering $\omega$ as
the planetary vorticity), this results in cyclones being pulled poleward
and anticyclones equatorward.

\subsubsection*{Meridional stability of circumpolar cyclones}

\begin{figure}
\begin{centering}
\includegraphics[width=1\columnwidth]{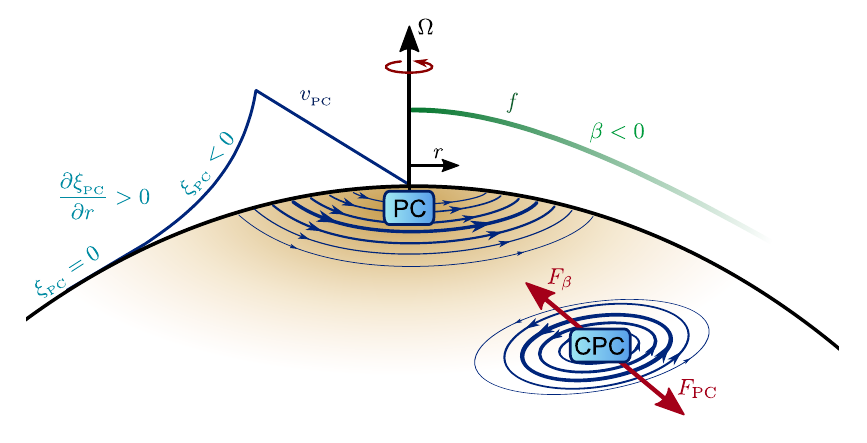}
\par\end{centering}
\caption{\textbf{An illustration of the balance holding a circumpolar cyclone
(CPC) around the polar cyclone (PC).} $r$ is the distance from the
pole (note that $r$ defines the positive meridional direction of
gradients). The green curve is the vertical component of the planetary
vorticity $f$. The blue curve is the profile around the PC of the
idealized axisymmetric tangential velocity in the vortex ($v_{{\rm PC}}$).
$\xi_{{\rm PC}}$ is the relative vorticity due to the presence of
the PC. $\Omega$ is the planetary rotation rate. The red arrows are
vorticity gradient forces on the CPC, induced by the PC and by planetary
sphericity.\label{fig: balance cartoon}}
\end{figure}
If the only vorticity gradient present in the background of a cyclone
is due to the planetary sphericity, it will tend to move poleward
until this gradient vanishes precisely at the pole. Assuming a PC
is already present at the pole (Fig.~\ref{fig: balance cartoon}),
the PC will induce a vorticity gradient of itself around the pole.
The velocity profiles observed for the PCs of Jupiter and Saturn resemble
a solid-disk rotation in an inner region, and an exponential decay
outside of it (Extended Data Figs.~1 and~2). Where velocity decays
exponentially, relative vorticity is negative due to the functional
relation between velocity and vorticity (Methods, Extended Data Fig.~3).
As the vorticity of the PC must vanish far away, an annulus of positive
relative vorticity gradient should exist around a PC due to its presence.
However, $f$ has its maximum magnitude at the poles and is $0$ at
the equator. Therefore, its gradient ($\beta\equiv\frac{\partial f}{\partial r}$)\textbf{
}is always negative away from the pole. The magnitude of $\beta$
is highest near the equator and vanishes at the poles\citep{vallis2017atmospheric}.
These trends mean that theoretically, there can be a latitude where
a poleward migrating CPC will be in equilibrium since the gradients
of vorticity due to the PC and due to the planetary sphericity are
equal and opposite\citep{li2020}. We propose this criterion of whether
a PC can generate a vorticity gradient that opposes $\beta$ as a
separating threshold between two polar states: one state in which
a circumpolar ring of vortices can be stably held, and one in which
the planetary gradient of vorticity is always greater, where any incoming
CPC will be merged into the PC.

Considering the meridional balance on a CPC, we define 

\begin{equation}
F_{\theta}\equiv\frac{\partial\xi_{{\rm PC}}}{\partial r}+\beta,\label{eq: Meridonal balance}
\end{equation}
which is proportional to the net meridional force on a CPC ($F_{{\rm PC}}-F_{\beta}$
in Fig.~\ref{fig: balance cartoon}). It is required that $F_{\theta}=0$
for a CPC to be in a meridional balance. Moreover, if $F_{\theta}$
is negative, the force is in the poleward direction; positive $F_{\theta}$
pushes cyclones equatorward.
\begin{figure*}[t]
\begin{centering}
\includegraphics[width=0.7\textwidth]{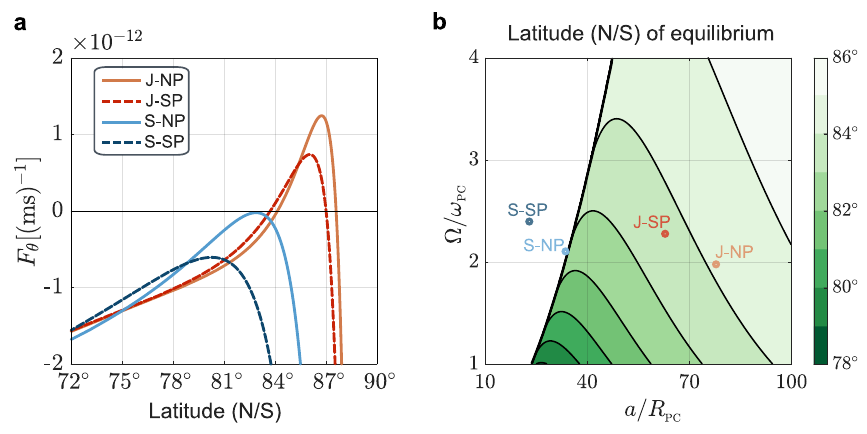}
\par\end{centering}
\caption{\textbf{Latitudes of equilibrium in the Gas Giants. a}, Curves of
$F_{\theta}$ as a function of latitude. The red curves are calculated
for the north (solid) and south (dashed) poles of Jupiter (J-NP and
J-SP, respectively); the blue curves represent the north (solid, S-NP)
and south (dashed, S-SP) poles of Saturn. Only Jupiter has points
of equilibrium in which $F_{\theta}$ is zero. A stable balance for
Jupiter's poles is achieved at the equilibrium points farther from
the pole. \textbf{b}, The latitude of equilibrium as a function of
the ratio between the planetary radius ($a$) and the radius of the
PC ($R_{{\rm PC}})$, and of the ratio between the planetary rotation
rate ($\Omega$) and the rotation rate of the PC ($\omega_{{\rm PC}}$).
Only the stable solutions for $F_{\theta}=0$ are considered. The
white area on the left side of the contour is where equilibrium can
not be achieved. The values representing the curves in \textbf{a}
are shown as points. Both poles of Saturn are in the region with no
solution. \label{fig: results}}
\end{figure*}
In Fig.~\ref{fig: results}a, $F_{\theta}$ is plotted as a function
of latitude. The four curves are drawn according to relative vorticity
gradients calculated from idealized profiles of the PCs' tangential
velocity (Methods). These profiles are determined according to the
maximum velocities and the radii of maximum velocity evaluated for
the respective PCs on the north and south poles of Jupiter and Saturn\citep{grassi2018first,baines2009saturn}
(Methods). Temporal and local variations of the velocity fields are
not taken into account. The $\beta$ profiles are calculated according
to the respective planetary radii and rotation rates of Jupiter and
Saturn (Methods).

It is evident (Fig.~\ref{fig: results}a) that both poles of Saturn
cannot sustain a meridional equilibrium, and therefore do not have
CPCs. In contrast, each pole of Jupiter exhibits two equilibrium points.
However, the equilibrium point closer to the pole (in each of the
red curves) is unstable. This is because a perturbation in the latitude
of the CPC poleward from that point will further pull it to merge
with the PC due to the negative vorticity gradient poleward of that
point. A perturbation away from the pole will bring the vortex to
the farther point of equilibrium. That point is in stable equilibrium,
situated at latitude $\sim84^{\circ}$ for both poles. The circumpolar
ring observed at J-NP lies approximately along $83^{\circ}$N, while
at J-SP, it is roughly at $84^{\circ}$S\citep{tabataba2020long}.
This agreement between the calculated latitudes of equilibrium (Fig.~\ref{fig: results})
and observations and the lack of such equilibrium on Saturn, support
the suggested mechanism as the stabilizing balance that holds the
CPCs of Jupiter stable.

It can be seen (Fig.~\ref{fig: results}b) that the stable equilibrium
is achieved farther from the poles when the PC rotates faster (smaller
$\Omega/\omega_{{\rm PC}}$). This is because the vorticity gradient
of the PC is proportional to its rotation rate, so faster rotating
PCs can ``overcome'' the planetary vorticity gradient for greater
distances from their centers. Larger polar vortices (smaller $a/R_{{\rm PC}}$)
also result in farther latitudes of equilibrium. In this case, this
is due to their vorticity gradient profiles being stretched farther.
This, however, has a limitation. A PC too big, relative to the planet,
will be in a state where its region of positive vorticity gradient
is too far from the pole, where $\beta$ dominates. In these cases,
equilibrium cannot be reached, as is the case of the Saturnian poles.

\subsubsection*{Zonal stability of circumpolar cyclones}

\begin{figure*}[!t]
\begin{centering}
\includegraphics[width=0.7\textwidth]{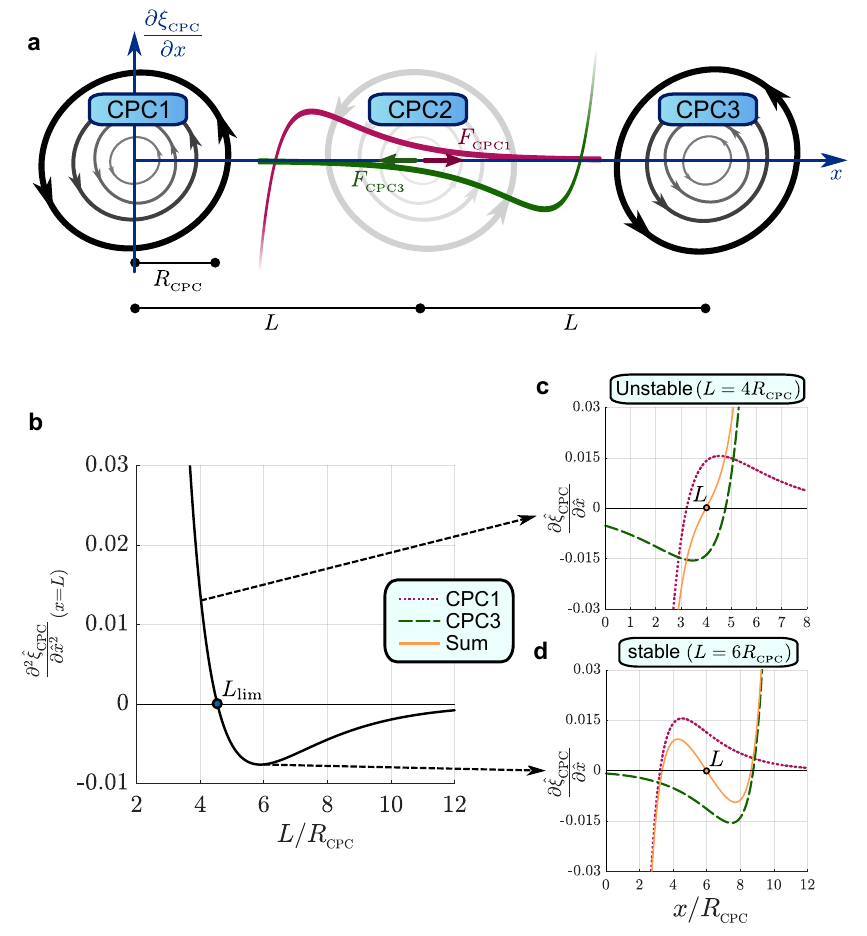}
\par\end{centering}
\caption{\textbf{Zonal stability of CPCs. a}, Schematic for the equilibrium
between CPC2 and the adjacent vortices in the zonal direction, titled
CPC1 and CPC3. $x$ is the distance from CPC1 in the zonal direction.
The radius of maximum velocity for a CPC is $R_{{\rm _{CPC}}}$. The
distance between two neighboring cyclones is $L$. $\xi_{{\rm CPC}}$
is the vorticity profile produced by the presence of CPCs 1 and 3.
The magenta and green curves are the vorticity gradients in the zonal
direction induced by CPC1 and CPC3, respectively. The rejection force
induced by each curve is illustrated as an arrow at the center of
CPC2. \textbf{b}, The second derivative of $\xi_{{\rm CPC}}$ in the
zonal direction, as a function of the distance $L$. This second derivative
is evaluated at a distance $L$ from CPC1.\textbf{ c},\textbf{d},
Plots representing an unstable and a stable case (respectively) of
the vorticity gradient profiles as a function of $x$.\textbf{ b}-\textbf{d},
For generalization of the plots, the hatted variables are non-dimensional
(Methods). \label{fig: zonal eq}}
\end{figure*}
Next, we investigate how many cyclones can fit in a circumpolar ring.
All cyclones in the circumpolar ring are of nearly similar size and
strength, with a comparable space between adjacent pairs\citep{tabataba2020long,adriani2020two,grassi2018first}.
Thus, in the following analysis, a ring of equally spaced identical
cyclones is considered. Then, it is postulated that the cyclones are
sustained by mutual rejection between their two neighboring vortices
in the zonal direction, according to the force derived by their respective
vorticity gradients (Fig.~\ref{fig: zonal eq}a). CPC2 is assumed
to be exactly in the middle between CPCs 1 and 3, such that the zonal
forces applied on it are equal and opposite. The remaining question,
however, is whether this equilibrium is stable. For understanding
the stability criterion for a ring of CPCs, it is insightful to consider
the second derivative in the zonal direction of the vorticity (only
the vorticity induced by CPCs 1 and 3) at the center of CPC2, as a
function of $L$ (Fig.~\ref{fig: zonal eq}b). When $L$ is too short,
a small perturbation in the location of CPC2 to the left encounters
a negative vorticity gradient (Fig.~\ref{fig: zonal eq}c), which
pulls the vortex further to the left. The opposite happens with a
perturbation to the right. This constitutes an unstable equilibrium
that will exist as long as the second derivative of the vorticity
is positive. When the second derivative of the vorticity is negative,
a stable equilibrium is formed (Fig.~\ref{fig: zonal eq}d). The
limiting distance, $L_{{\rm lim}}=4.54R_{{\rm CPC}}$ (Methods), in
which the second derivative vanishes, is the minimal distance between
CPCs that can maintain a stable equilibrium. This means that the vortex
centers have to be more than $L_{{\rm lim}}$ apart in order for the
CPCs to be in a sustainable configuration.

The available space for CPCs in the circumferential ring is approximately
$2\pi a(90^{\circ}-\left|\theta_{{\rm eq}}\right|)\pi/180^{\circ}$,
where $\theta_{{\rm eq}}$ is the latitude of equilibrium for the
respective pole (Fig.~\ref{fig: results}b). Therefore, the maximal
number of vortices to fit in the ring can be estimated according to
\begin{equation}
N_{{\rm max}}=\frac{2\pi a(90^{\circ}-\left|\theta_{{\rm eq}}\right|)\pi/180^{\circ}}{4.54R_{{\rm _{CPC}}}}.\label{eq: Nmax equation}
\end{equation}
Here, values for $R_{{\rm _{CPC}}}$ at the north and south poles
of Jupiter are evaluated by Juno's JIRAM imager\citep{adriani2020two}.
Higher values inside the variability range of $R_{{\rm _{CPC}}}$
were taken for the smallest constraint (Methods). Inserting the numbers
for the north and south poles to equation~(\ref{eq: Nmax equation})
results in $N\sim11.05$ and $\sim7.26$, respectively. However, as
$N$ should be an integer describing the maximal number of stable
CPCs, these numbers are rounded down to give

\begin{equation}
\begin{aligned}N_{{\rm max,N}}\sim & 11, & N_{{\rm max,S}}\sim7\end{aligned}
,
\end{equation}
where the subscripts N and S represent the north and south poles.

While this last analysis can be explained intuitively, to get a more
accurate constraint, a 2D analysis is performed that does not treat
CPC2 as a singular point but instead considers the different influence,
weighted by the meridional velocity of CPC2, around the spread of
the vortex. The force acting on CPC2 by the presence of CPCs 1 and
3 is proportional\citep{rossby1948displacements} to the integral
\begin{equation}
F\sim\iintop v_{{\rm 2}}(\xi_{{\rm 1}}+\xi_{3})dS\label{eq: Rossbey integral}
\end{equation}
around CPC2. Here, $v_{{\rm 2}}$ is the meridional velocity of CPC2,
and $\xi_{{\rm 1(3)}}$is the vorticity of CPC1(3). The results from
this analysis are qualitatively similar to those shown in Fig.~\ref{fig: zonal eq};
however, the limiting distance for stability ($L_{{\rm lim}}$) is
found to be $5.87R_{{\rm CPC}}$ instead (Methods). This constrains
$N_{{\rm max}}$ further as CPC2 is now influenced more by regions
closer to CPCs 1 and 3. Using this value instead of $4.54R_{{\rm _{CPC}}}$
in equation~(\ref{eq: Nmax equation}) results in $N\sim8.54$ and
$\sim5.62$ for the north and south poles, such that
\begin{equation}
\begin{aligned}N_{{\rm max,N}}\sim8 & , & N_{{\rm max,S}}\sim5\end{aligned}
.
\end{equation}
This constraint correctly predicts the actual number of vortices in
the north and south poles of Jupiter. It is interesting to note that
the estimation of $N_{{\rm max,S}}$ is in between $5$ and $6$,
and, consistently, two years observations of the south pole found
a constant gap in the south polar ring\citep{tabataba2020long}, implying
on space that is slightly larger than that required for the five CPCs.
This gap was temporarily occupied with a sixth vortex around the time
of Juno's 18th perijove\citep{adriani2020two}, but as the formation
was not stable, this additional vortex disappeared by the time of
the 19th perijove, indicating that the 5.62 value has a dynamical
meaning. These predictions further support the described mechanism
for the stability of the CPCs on Jupiter.

\subsubsection*{Polar cyclones on the Gas Giants \label{sec:Discussion}}

The analysis presented here is based on the assumption that large-scale
movement of vortices is mainly due to advection of background vorticity
with the tangential velocity of the vortex, and that this movement
is proportional, both in magnitude and in direction, to the background
vorticity gradient. This logic implies that if the background vorticity
gradient is zero at the center of a vortex, that vortex will not move.
The gradient of planetary vorticity that acts on a circumpolar cyclone
in the meridional direction can be opposed under certain conditions
by the gradient of vorticity induced by the polar cyclone (equation~(\ref{eq: Meridonal balance})).
Such an equilibrium is shown (Fig.~\ref{fig: results}b) to be favored
for small and strong PCs, relative to their host planets. In the zonal
direction, it is shown (Fig.~\ref{fig: zonal eq}) that stability
can be sustained for up to a certain amount of vortices in a circumpolar
ring. More vortices can fit in the ring for poles in which the meridional
stability is achieved at latitudes farther from the pole, and for
poles with smaller CPCs. These analyses only treat the assumed highest-order
forces that control the stability of circumpolar vortices, and suggest
that the governing dynamics are controlled by 2D (barotropic) PV conservation.
There are other forces, such as ones deriving from a 3D structure
of a cyclone, that may be responsible for the small changes in the
locations, speeds, and sizes observed in the circumpolar vortices
of Jupiter\citep{tabataba2020long}.

While this study explains the existence of circumpolar cyclones on
Jupiter in contrast to the absence of such on Saturn, it does so by
considering the specifics of their observed corresponding PCs. However,
it does not explain the variation between the PCs of the planets.
Theoretically, Saturn could sustain circumpolar vortices if its PCs
were smaller or were to spin faster. Since the stability criterion
for a circumpolar ring in the North Pole of Saturn is nearly reached,
it may be that minor future variations in the polar conditions of
Saturn (e.g., due to its seasonality) would manifest in a circumpolar
ring or that such a ring existed in the past. It is also possible
that cyclic variations in the solar forcing may alternate the number
of circumpolar vortices between the north and south poles of Jupiter.
Nonetheless, the match of the meridional and zonal force balances
to the observations provides strong evidence that the physical balances
outlined in this study are responsible for setting the location, stability,
and number of circumpolar cyclones on the Gas Giants.

\section*{Acknowledgments}

We thank Keren Duer and Eli Galanti for insightful conversations.
This research has been supported by the Minerva Foundation with funding
from the Federal German Ministry for Education and Research and the
Helen Kimmel Center for Planetary Science at the Weizmann Institute
of Science.

\section*{Author contributions}

N.G. has designed the study, performed the calculations and written
the paper with support of Y.K.

\section*{Additional information}

\paragraph*{Correspondence and requests for materials}

should be addressed to N.G. (nimrod.gavriel@weizmann.ac.il)

\bibliographystyle{naturemag}
\bibliography{Nimrodbib}

\appendix
\onecolumn

\section*{Methods}

\paragraph*{Idealized cyclone profiles.}

The observed velocity profiles for the vortices at the poles of Jupiter\citep{grassi2018first}
are very similar to a solid-disk rotation in an inner region and an
exponential decay outside of it (Extended Data Fig.~1). This behavior
is expressed by an idealized cyclone tangential velocity profile given
as 
\begin{equation}
v_{{\rm PC}}=\begin{cases}
V\frac{r}{R} & 0\leq r<R\\
Ve^{-\frac{r-R}{2R}} & r\geq R
\end{cases},\label{eq: Piecewise velocity}
\end{equation}
where $r$ is the distance from the center of the cyclone, $V$ is
the maximum velocity in the vortex, and $R$ is the edge of the solid
disk. This profile is compared in Extended Data Fig.~3a with a profile
suggested for a study of tropical cyclones\citep{smith1990numerical}
and with a profile fit specifically for the PCs of Jupiter\citep{grassi2018first}.
This velocity profile (equation~(\ref{eq: Piecewise velocity}))
is also compared with wind measurements from Jupiter and Saturn in
Extended Data Figs.~1 and~2, respectively. The stronger decay of
velocity around $3,000$~km at the Jovian poles (Extended Data Fig.~1)
can be attributed to the presence of the velocity fields of the CPCs,
as no such trend appears in the PC only cases of the Saturnian poles
(Extended Data Fig.~2).\textbf{ }The relative vorticity around the
center of the cyclone is calculated for a cyclone put in a medium
otherwise at rest as $\xi=\frac{1}{r}\frac{\partial\left(rv_{{\rm PC}}\right)}{\partial r}$
to give 
\begin{equation}
\xi_{{\rm PC}}=\begin{cases}
2\frac{V}{R} & 0\leq r<R\\
\left(\frac{R}{r}-\frac{1}{2}\right)\frac{V}{R}e^{-\frac{r-R}{2R}} & r\geq R
\end{cases}.\label{eq: Piecewise vorticity}
\end{equation}
This profile is compared with the vorticity calculated from the two
other velocity profiles in Extended Data Fig.~3b. The vorticity gradient
is thus 
\begin{equation}
\frac{\partial\xi_{{\rm PC}}}{\partial r}=\begin{cases}
0 & 0\leq r<R\\
\frac{V}{R^{2}}e^{-\frac{r-R}{2R}}\left(\frac{1}{4}-\frac{R}{2r}-\frac{R^{2}}{r^{2}}\right) & r\geq R
\end{cases}.\label{eq: Vortex vorticity gradient}
\end{equation}
This profile is again compared with the two other profiles in Extended
Data Fig.~3c.\textbf{ }The velocity profile from the numerical study\citep{smith1990numerical}\textit{
}is an inverse high order polynomial. This means that the second derivative
of this velocity profile, which is the requested term, is very noisy.
On the other hand, the curve fit for J-NP\citep{grassi2018first}
is only suited for a small range of $r$ and shows a large vorticity
gradient near $r=R$ that is an artifact of the chosen curve. For
these reasons, we chose to perform the calculations of this study
with the suggested piece-wise profile (equations~(\ref{eq: Piecewise velocity})-(\ref{eq: Vortex vorticity gradient})).

\paragraph*{Equations for the meridional stability}

The planetary background vorticity ($f$) is

\begin{equation}
f=2\Omega\cos(r/a),
\end{equation}
where $\Omega$ is the planetary rotation rate, and $a$ is the radius
of the planet. The planetary vorticity gradient is 
\begin{equation}
\beta\equiv\frac{\partial f}{\partial r}=-2\Omega a^{-1}\sin(r/a).\label{eq:beta definition}
\end{equation}
In order to show Fig.~\ref{fig: results} in terms of latitude, the
transformation $r=a(90^{\circ}-\theta)\pi/180^{\circ}$ is used in
equation~(\ref{eq: Meridonal balance}). The two terms in equation~(\ref{eq: Meridonal balance})
are defined by equation~(\ref{eq: Vortex vorticity gradient}) and
equation~(\ref{eq:beta definition}). For the planetary rotation
rates ($\Omega$), the values $1.76\times10^{-4}$ and $1.65\times10^{-4}$
$\thinspace\text{s}^{-1}$ are used for Jupiter and Saturn, respectively.
The mean planetary radii ($a$) used here are $69,911$\,km for Jupiter
and $58,232$\,km for Saturn. Maximum velocities ($V$) and radii
of maximum velocity ($R=R_{{\rm PC}}$) are estimated for Jupiter
(Grassi et al., 2018\citep{grassi2018first}, Fig.~6) and Saturn
(Baines et al., 2009\citep{baines2009saturn}, Tab.~2) from observations.
For J-NP, J-SP, S-NP, and S-SP, those values are $V=\left\{ 80,85,136,174\right\} {\rm \thinspace ms^{-1}}$
and $R_{{\rm PC}}=\left\{ 900,1100,1728,2541\right\} {\rm \thinspace km}$,
respectively.

In Fig.~\ref{fig: results}b, the contour shows the solutions $\theta_{{\rm eq}}$
for $F_{\theta}=0$, reduced to

\begin{equation}
\theta_{{\rm eq}}=G(a/R_{{\rm PC}},\Omega/\omega_{{\rm PC}}),
\end{equation}
where $G$ is a function defined by using equation~(\ref{eq: Vortex vorticity gradient})
and equation~(\ref{eq:beta definition}), and $\omega_{{\rm PC}}=V/R_{{\rm PC}}$
is the rotation rate of the solid-disk part of the PC. While at some
range of $a/R_{{\rm PC}}$ and $\Omega/\omega_{{\rm PC}}$ there is
no solution, as can be seen in Fig.~\ref{fig: results}b, for the
rest of the range, there are two possible solutions. Only the stable
solutions (where $\frac{\partial F_{\theta}}{\partial\theta}\Bigr|_{\theta=\theta_{{\rm eq}}}>0$)
are taken for Fig.~\ref{fig: results}b.

\paragraph*{Equations for zonal stability}

For plotting the general trends in Fig.~\ref{fig: zonal eq}, equation~(\ref{eq: Vortex vorticity gradient})
(for the CPC, in the zonal direction) is normalized according to 
\begin{equation}
r=\hat{x}R_{{\rm CPC}},\quad\xi=\hat{\xi}V_{{\rm CPC}}/R_{{\rm CPC}},\label{eq: scaling}
\end{equation}
where variables with a circumflex are non-dimensional, $\hat{x}$
is the non-dimensional distance from the center of CPC1 in the eastward
direction, $R_{{\rm CPC}}$ is the radius of maximum velocity of the
CPCs, and $V_{{\rm CPC}}$ is the maximum velocity of the CPCs. These
scalings result in equation (\ref{eq: Vortex vorticity gradient})
becoming
\begin{equation}
\frac{\partial\hat{\xi}}{\partial\hat{x}}=\begin{cases}
0 & 0\leq\hat{x}<1\\
e^{-\frac{\hat{x}-1}{2}}\left(\frac{1}{4}-\frac{1}{2\hat{x}}-\frac{1}{\hat{x}^{2}}\right) & \hat{x}\geq1
\end{cases}.
\end{equation}
The total vorticity gradient in the zonal direction, felt on CPC2
by CPCs 1 and 3, is therefore 
\begin{equation}
\frac{\partial\hat{\xi}_{{\rm CPC2}}}{\partial\hat{x}}=\frac{\partial\hat{\xi}}{\partial\hat{x}}-\left[\frac{\partial\hat{\xi}}{\partial\hat{x}}\right]_{\hat{x}\rightarrow\left(2L/R_{{\rm CPC}}-\hat{x}\right)}.\label{eq: zonal eq}
\end{equation}
In Fig.~\ref{fig: zonal eq}b, the expression $\left[\frac{\partial}{\partial\hat{x}}\left(\frac{\partial\hat{\xi}_{{\rm CPC2}}}{\partial\hat{x}}\right)\right]_{\hat{x}\rightarrow L/R_{{\rm CPC}}}$
is plotted against $L/R_{{\rm CPC}}$. The minimum distance between
CPCs required for stability ($L_{{\rm lim}}$) is the solution $L$
for the equation
\begin{equation}
\left[\frac{\partial}{\partial\hat{x}}\left(\frac{\partial\hat{\xi}_{{\rm CPC2}}}{\partial\hat{x}}\right)\right]_{\hat{x}\rightarrow L/R_{{\rm CPC}}}=0.
\end{equation}
This value is found to be $L_{{\rm lim}}=4.54R_{{\rm CPC}}$. In Fig.~\ref{fig: zonal eq}c-d,
equation~(\ref{eq: zonal eq}) is plotted against $\hat{x}$ for
two different values of $L$. In Fig.~\ref{fig: zonal eq}c, $L$
is smaller than $L_{{\rm lim}}$, illustrating an unstable equilibrium
that would result in a merger with either CPC1 or 3. In Fig.~\ref{fig: zonal eq}d,
$L$ is larger than $L_{{\rm lim}}$, and the equilibrium is stable.

For solving equation~(\ref{eq: Nmax equation}), $R_{{\rm CPC}}$
is evaluated according to $R_{{\rm CPC}}=R_{{\rm PC}}R_{{\rm ratio}}^{-1}$
, where the ratios between the radii of the PC and the CPCs ($R_{{\rm ratio}}$)
are estimated from Adriani et al. 2020\citep{adriani2020two} (Fig.~5b).
We used the minimal (for most restrictive constraint) observed values
of $R_{{\rm ratio}}=1$ for J-NP, and $R_{{\rm ratio}}=0.75$ for
J-SP. Values for $R_{{\rm PC}}$ are the same as for the meridional
analysis. For $\left|\theta_{{\rm eq}}\right|$, the values $84.1^{\circ}$
and $83.7^{\circ}$ found from Fig.~\ref{fig: results} for J-NP
and J-SP, respectively, were used. The resulting $N_{{\rm max}}$
was rounded down, as rounding up would result in an unstable amount
of CPCs.

\paragraph*{Accurate estimation of $L_{{\rm lim}}$}

For a more accurate prediction of the limiting distance between vortices
for stability, a 2D analysis is done instead of the 1D analysis done
for Fig.~\ref{fig: zonal eq}. We use here the notion that the vorticity
gradient force is proportional to the integral shown in equation~(\ref{eq: Rossbey integral}).
Stability is achieved when this force is positive (pushes right) when
the position of CPC2 is perturbed to the left and is negative when
this position is perturbed to the right. Therefore, the limiting $L$
for stability ($L_{{\rm lim}}$) is the value of $L$ in which the
gradient of the force with respect to the location of CPC2 vanishes.

For performing the integration in equation~(\ref{eq: Rossbey integral}),
the ideal profiles (equations~(\ref{eq: Piecewise velocity}) and
(\ref{eq: Piecewise vorticity})) are converted to a Cartesian coordinate
system. This gives (in normalized variables according to equation~(\ref{eq: scaling}),
where $v=\hat{v}V_{{\rm CPC}}$) 
\begin{equation}
\begin{alignedat}{2}\hat{v}_{2}= & \begin{cases}
r_{2}\cos(\phi)\\
e^{-\frac{r_{2}-1}{2}}\cos(\phi)
\end{cases} & \begin{array}{c}
0\leq r_{2}<1\\
r_{2}\geq1
\end{array}\text{,}\\
\hat{\xi}_{1}= & \begin{cases}
2\\
\left(\frac{1}{r_{1}}-\frac{1}{2}\right)e^{-\frac{r_{1}-1}{2}}
\end{cases} & \begin{array}{c}
0\leq r_{1}<1\\
r_{1}\geq1
\end{array},\\
\hat{\xi}_{3}= & \begin{cases}
2\\
\left(\frac{1}{r_{3}}-\frac{1}{2}\right)e^{-\frac{r_{3}-1}{2}}
\end{cases} & \begin{array}{c}
0\leq r_{3}<1\\
r_{3}\geq1
\end{array},
\end{alignedat}
\end{equation}
where 
\begin{equation}
\begin{alignedat}{1}\hat{r}_{1}= & \sqrt{\hat{x}^{2}+\hat{y}^{2}},\\
\hat{r}_{2}= & \sqrt{\left(\hat{x}-\hat{x}_{{\rm 2}}\right)^{2}+\hat{y}^{2}},\\
\hat{r}_{3}= & \sqrt{\left(\hat{x}-2L/R_{{\rm CPC}}\right)^{2}+\hat{y}^{2}},\\
\phi= & \tan^{-1}\left(\frac{\hat{y}}{\hat{x}-\hat{x}_{{\rm 2}}}\right),
\end{alignedat}
\end{equation}
$\hat{x}_{2}$ is the normalized (by $R_{{\rm CPC}}$) distance between
the centers of CPC1 and CPC2, and $\hat{y}$ is the northward meridional
distance from the center of CPC1, normalized by $R_{{\rm CPC}}$.
$L_{{\rm lim}}$ is then the value of $L$ that solves the equation
\begin{equation}
\left[\frac{\partial}{\partial\hat{x}_{2}}\left(\iintop\hat{v}_{{\rm 2}}(\hat{\xi}_{{\rm 1}}+\hat{\xi}_{3})dS\right)\right]_{\hat{x}_{2}=L/R_{{\rm CPC}}}=0.\label{eq: 2D Llim eq}
\end{equation}
For validation, the domain of integration is determined to be very
small, in which case the resulting $L_{{\rm lim}}$ approached the
value from the 1D analysis ($4.54R_{{\rm CPC}}$). The domain of integration
is ultimately chosen to be $\left\{ \hat{x}_{2}-\left(L/R_{{\rm CPC}}-2\right),\hat{x}_{2}+\left(L/R_{{\rm CPC}}-2\right)\right\} $
for $\hat{x}$, and $\left\{ -\left(L/R_{{\rm CPC}}-2\right),\left(L/R_{{\rm CPC}}-2\right)\right\} $
for $\hat{y}$. This way, the resulting force is not stemming from
the predominant areas of CPCs 1 and 3. The solution $L_{{\rm lim}}$
to equation (\ref{eq: 2D Llim eq}) is thus $L_{{\rm lim}}=5.87$.

\paragraph*{Formal asymptotic derivation of the suggested balance on the CPCs:
momentum balance approach}

In order to show how the described balance suggested in this study
results from the equations of motion, an asymptotic derivation is
laid out. First, a time scale that is long enough to describe the
changes in the CPCs is needed. This time-scale is derived here from
the vorticity equation by balancing between the vorticity change with
time and the $\beta$ term, as this term is assumed to be a substantial
contributor to the motion of Jovian cyclones. Thus, it follows that

\begin{equation}
\frac{\partial\xi}{\partial t}\propto\beta v,
\end{equation}
leading to the scaling argument
\begin{equation}
T\propto\frac{1}{\beta L}\approx11{\rm \left[days\right]},
\end{equation}
where Jovian values were assumed, $\beta$ was estimated at latitude
$84^{\circ}$ (near the center of the CPCs), and $L\approx2000$ km
was estimated according to the observed radii of the CPCs. This time
scale is of the same order as the time between perijoves (\textasciitilde 53
days), in which slight variations can be observed in the locations
and sizes of the CPCs\citep{adriani2020two,tabataba2020long}. To
continue, we assume that the flow is 2D, inviscid, and barotropic.
We start from the horizontal conservation of momentum equation\citep{vallis2017atmospheric}
in the form 
\begin{equation}
\frac{\partial{\bf u}}{\partial t}+\left(\boldsymbol{f}+\boldsymbol{\xi}\right){\rm \boldsymbol{\hat{k}}}\times{\bf u}=-\frac{1}{\rho}\nabla p-\frac{1}{2}\nabla\left({\bf u}^{2}\right),
\end{equation}
where $p$ is pressure and ${\bf u}$ is the velocity vector. We scale
and expand the variables at the core of a CPC according to 
\begin{equation}
\begin{aligned}\begin{array}{c}
x=L\hat{x},\;\;\;\;\;y=L\hat{y},\;\;\;t=\left(\beta L\right)^{-1}\hat{t},\;\;\;\;\;u=U\left(\hat{u}_{{\rm CPC}}+{\rm Ro\,}\hat{u}_{{\rm PC}}\right),\;\;\;\;\;v=U\left(\hat{v}_{{\rm CPC}}+{\rm Ro\,}\hat{v}_{{\rm PC}}\right),\\
p=f_{0}UL\rho\left(\hat{p}_{{\rm CPC}}+{\rm Ro}\,\hat{p}_{{\rm PC}}\right),\;\;\;\;\;f=(f_{0}+\beta L\hat{y}),\;\;\;\;\;\xi=\frac{U}{L}\left(\hat{\xi}_{{\rm CPC}}+{\rm Ro}\,\hat{\xi}_{{\rm PC}}\right).
\end{array}\end{aligned}
\end{equation}
Here, ${\rm Ro}$ is the Rossby number, taken as the small asymptotic
expansion constant (${\rm Ro}=\frac{U}{f_{0}L}\approx0.05$), where
$U$ is the velocity scale ($\sim40\,{\rm ms^{-1}})$, and $f_{0}$
is evaluated at latitude $84^{\circ}$. The velocities (and therefore
the pressure and vorticity as well) due to the PCs are assumed to
be $O({\rm Ro})$ smaller than those of the CPC since the PC is far
when looking at the core of a CPC. The pressure here is scaled according
to geostrophic balance. Note that here the meridional direction is
northward ($y)$ and not equatorward ($r$) as defined in the main
text. This expansion results in
\begin{equation}
\begin{array}{c}
\frac{\beta L}{f_{0}}\left(\frac{\partial{\bf \hat{u}}_{{\rm CPC}}}{\partial\hat{t}}\right)+\left(1+\frac{\beta L}{f_{0}}\hat{y}+{\rm Ro}\left(\hat{\xi}_{{\rm CPC}}+{\rm Ro}\,\hat{\xi}_{{\rm PC}}\right)\right){\rm \boldsymbol{\hat{k}}}\times\left(\hat{{\bf u}}_{{\rm CPC}}+{\rm Ro}\hat{{\bf u}}_{{\rm PC}}\right)\\
\;\;\;\;\;\;\;\;\;\;\;\;\;\;\;\;\;\;\;\;=-\nabla\left(\hat{p}_{{\rm CPC}}+{\rm Ro}\,\hat{p}_{{\rm PC}}\right)-{\rm Ro}\frac{1}{2}\nabla\left({\bf \hat{u}}_{{\rm CPC}}\cdot\hat{{\bf u}}_{{\rm CPC}}+2{\rm Ro}\hat{{\bf u}}_{{\rm PC}}\cdot{\bf \hat{u}}_{{\rm CPC}}+O({\rm Ro^{2}})\right),
\end{array}
\end{equation}
where $\hat{k}$ is a unit vector in the vertical direction. The non-dimensional
number $\frac{\beta L}{f_{0}}$ is evaluated at latitude $84^{\circ}$
as $\sim0.003$, which is $O({\rm Ro^{2}})$. In the leading order
we have (back in dimensional variables) a geostrophic balance on the
CPC
\begin{equation}
f_{0}{\rm \boldsymbol{\hat{k}}}\times{\bf u}_{{\rm CPC}}=-\frac{1}{\rho}\nabla p_{{\rm CPC}}.\label{eq: Geostrophic CPC}
\end{equation}
For the second order we get a modified geostrophic balance for the
PC 
\begin{equation}
f_{0}{\rm \boldsymbol{\hat{k}}}\times{\bf u}_{{\rm PC}}=-\frac{1}{\rho}\nabla p_{{\rm PC}}-\xi_{{\rm CPC}}{\rm \boldsymbol{\hat{k}}}\times{\bf u}_{{\rm CPC}}-\nabla\left({\bf u}_{{\rm CPC}}\cdot{\bf u}_{{\rm CPC}}\right).\label{eq: Geostrophic PC}
\end{equation}
For the third order, where the time evolution appears, we have 
\begin{equation}
\frac{\partial{\bf {\bf u}_{{\rm CPC}}}}{\partial t}+\left(\beta y+\xi_{{\rm PC}}\right){\rm \boldsymbol{\hat{k}}}\times{\bf u}_{{\rm CPC}}+\xi_{{\rm CPC}}{\rm \boldsymbol{\hat{k}}}\times{\bf u}_{{\rm PC}}=-\nabla\left({\bf u}_{{\rm PC}}\cdot{\bf u}_{{\rm CPC}}\right).\label{eq: Momentum third order}
\end{equation}
Expanding equation~(\ref{eq: Momentum third order}) in the meridional
direction, while using equation~(\ref{eq: Geostrophic CPC}), gives
\begin{equation}
\frac{\partial v_{{\rm CPC}}}{\partial t}+u_{{\rm CPC}}\left(\xi_{{\rm PC}}+\beta y\right)=-\frac{1}{f_{0}\rho}\frac{\partial}{\partial y}\left(\frac{\partial p_{{\rm CPC}}}{\partial x}u_{{\rm PC}}\right)-\text{\ensuremath{v_{{\rm CPC}}}}\frac{\partial v{}_{{\rm PC}}}{\partial y}-v_{{\rm PC}}\frac{\partial v{}_{{\rm CPC}}}{\partial y}.\label{eq: meridional 3rd order momentum equation}
\end{equation}
Since the PC terms are zonally symmetric, and since the CPC terms
on the right-hand side are anti-symmetric in the zonal direction relative
to the core of the CPC, a concentric integration of equation~(\ref{eq: meridional 3rd order momentum equation})
around the core of the CPC results in the vanishing of all the terms
on the right-hand side. Therefore, we get that 
\begin{equation}
F_{y}=-\iint u_{{\rm CPC}}\left(\xi_{{\rm PC}}+\beta y\right)dS,
\end{equation}
where $F_{y}$ is the meridional force density on the CPC core. By
considering the anti-symmetry of $u_{{\rm CPC}}$ in the meridional
direction, one can show that $F_{y}$ vanishes only when $\iint\frac{\partial}{\partial y}\left(\xi_{{\rm PC}}+\beta y\right)dS=0$,
resulting in the condition $F_{\theta}=0$ (equation~(\ref{eq: Meridonal balance})).
Taking the zonal direction of equation~(\ref{eq: Momentum third order}),
and replacing the velocities according to Fig.~\ref{fig: zonal eq}
would similarly result in 
\begin{equation}
F_{x}=\iint v_{{\rm CPC}2}\left(\xi_{{\rm CPC1}}+\xi_{{\rm CPC3}}\right)dS.
\end{equation}

\paragraph*{Formal asymptotic derivation of the suggested balance on the CPCs:
vorticity balance approach}

Another way to arrive at the condition $F_{\theta}=0$ (equation~(\ref{eq: Meridonal balance}))
is to look at the vorticity equation. Taking the curl of equation~(\ref{eq: Geostrophic CPC})
gives
\begin{equation}
\nabla\cdot{\bf u}_{{\rm CPC}}=0.\label{eq: Ucpc non divergent}
\end{equation}
 Taking the curl of equation~(\ref{eq: Geostrophic PC}) gives
\begin{equation}
\left({\bf u}_{{\rm CPC}}\cdot\nabla\right)\xi_{{\rm CPC}}+f_{0}\left(\nabla\cdot{\bf u}_{{\rm PC}}\right)=0.
\end{equation}
 Assuming that the core of the CPC rotates, in equilibrium, as a solid
body ($\frac{\partial\xi_{{\rm CPC}}}{\partial x}=\frac{\partial\xi_{{\rm CPC}}}{\partial y}=0$)
gives that 
\begin{equation}
\nabla\cdot{\bf u}_{{\rm PC}}=0.\label{eq: Upc Nondivergent}
\end{equation}
 Taking the curl of equation~(\ref{eq: Momentum third order}) gives
\begin{equation}
\text{\ensuremath{\frac{\partial\xi_{{\rm CPC}}}{\partial t}}}+v_{{\rm CPC}}\left(\text{\ensuremath{\frac{\partial\xi_{{\rm PC}}}{\partial y}}}+\beta\right)+(\xi_{{\rm PC}}+\beta y)\left(\nabla\cdot{\bf u}_{{\rm CPC}}\right)+\xi_{{\rm CPC}}\left(\nabla\cdot{\bf u}_{{\rm PC}}\right)+\left({\bf u}_{{\rm PC}}\cdot\nabla\right)\xi_{{\rm CPC}}+\text{\ensuremath{\frac{\partial\xi_{{\rm PC}}}{\partial x}}}u_{{\rm CPC}}=0.
\end{equation}
Using the solid-body rotation assumption again, together with equations~(\ref{eq: Ucpc non divergent})
and (\ref{eq: Upc Nondivergent}), and noting that $\xi_{{\rm PC}}$
is zonally symmetric gives 
\begin{equation}
\text{\ensuremath{\frac{\partial\xi_{{\rm CPC}}}{\partial t}}}=-v_{{\rm CPC}}\left(\text{\ensuremath{\frac{\partial\xi_{{\rm PC}}}{\partial y}}}+\beta\right),
\end{equation}
where the term in the parentheses can be regarded as $\frac{\partial\omega}{\partial y}$
in Fig.~\ref{fig: beta-drift-fig}. As $v_{{\rm CPC}}$ is anti-symmetric
in the zonal direction, two opposite vorticity anomalies can be generated
in the two sides of the CPC core when $\frac{\partial\omega}{\partial y}\neq0$,
resulting in a net meridional acceleration on the core.

\section*{Data Availability}

No data sets were generated or analyzed during the current study.

\section*{Code availability}

The MATLAB codes used for calculating and plotting the figures in
this paper are available on request from N.G.

\section*{Competing interests}

The authors declare no competing interests.

\renewcommand{\thefigure}{{\arabic{figure}}}
\setcounter{figure}{0}     
\renewcommand{\figurename}{{Extended Data Fig.}} 

\begin{figure}[H]
\begin{centering}
\includegraphics[width=1\columnwidth]{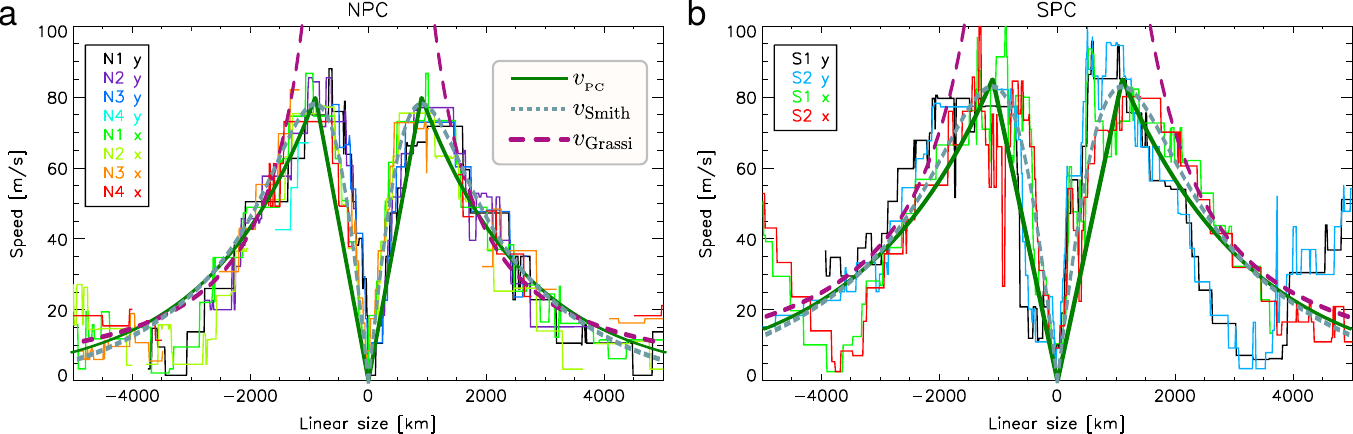}
\par\end{centering}
\centering{}\caption{\textbf{Measurements of the Jovian PC velocity profiles}. The velocity
profiles from Extended Data Fig.~3a, overlaid on Fig.~6 from Grassi
et al., 2018\citep{grassi2018first} (adapted with permission), showing
the observed velocities around the north (\textbf{a}) and south (\textbf{b})
poles of Jupiter. The idealized velocity profiles were calculated
using the Jovian values for $R$ and $V$ (Methods). The green curves
($v_{{\rm PC}}$) represent the velocity profiles used for the analyses
in this study.\label{fig: Jupiter polar velocities}}
\end{figure}
\begin{figure}[H]
\begin{centering}
\includegraphics[width=0.6\columnwidth]{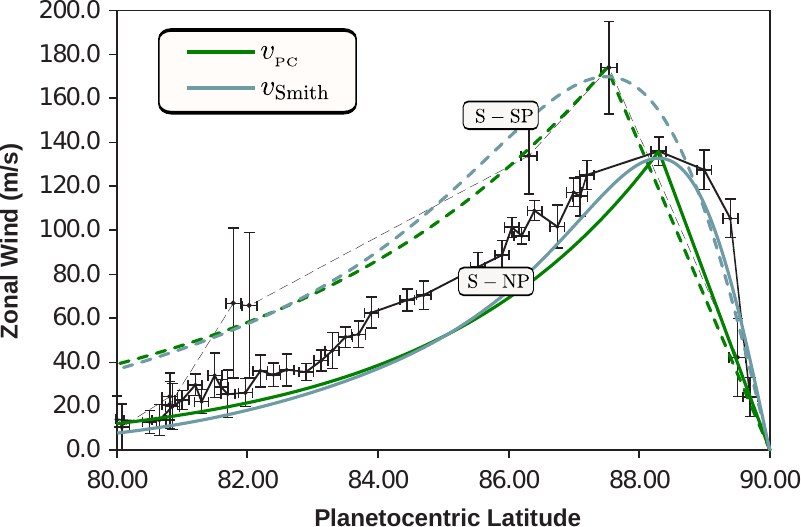}
\par\end{centering}
\centering{}\caption{\textbf{Measurements of the Saturnian PC velocity profiles}. Two velocity
profiles from Extended Data Fig.~3a, overlaid on Fig.~8 from Baines
et al., 2009\citep{baines2009saturn} (adapted with permission), showing
the observed velocities around the north (solid) and south (dashed)
poles of Saturn. Error bars are calculated as standard deviations\citep{baines2009saturn}.
The idealized velocity profiles were calculated using the Saturnian
values for $R$ and $V$ (Methods). The green curves ($v_{{\rm PC}}$)
represent the velocity profiles used for the analyses in this study.\label{fig: Saturn polar velocities}}
\end{figure}
\begin{figure}[H]
\begin{centering}
\includegraphics[width=1\columnwidth]{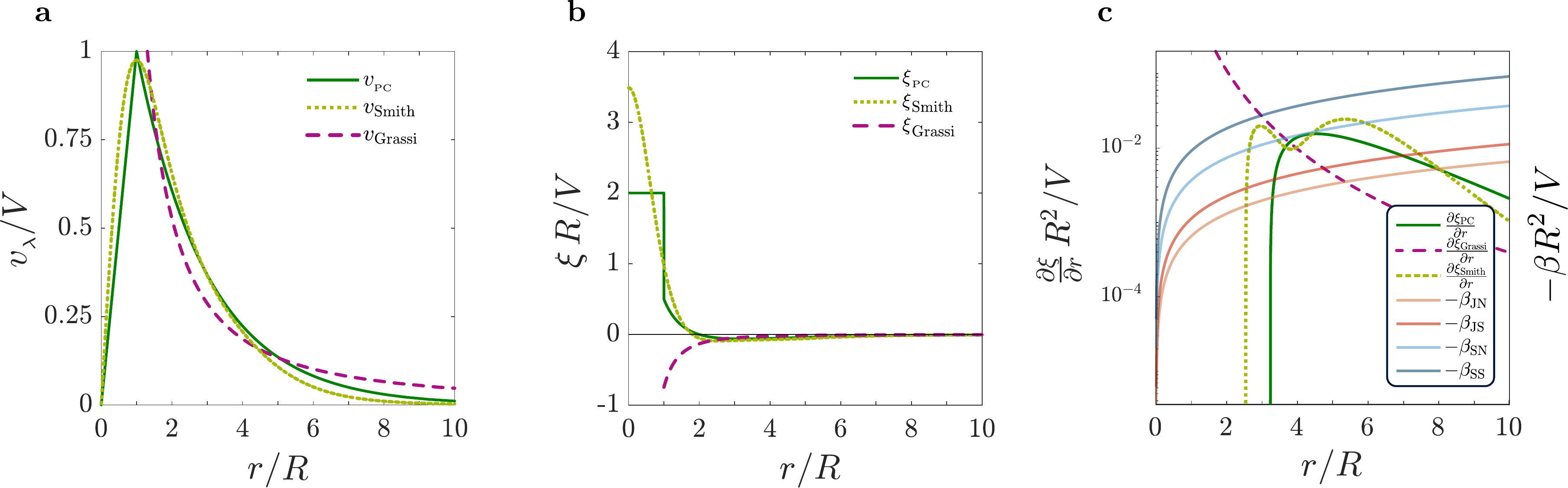}
\par\end{centering}
\centering{}\caption{\textbf{Idealized profiles of velocity, vorticity, and vorticity gradient.}
\textbf{a},\textbf{ }The vortex velocity profile according to the
suggested piece-wise function (green solid curve) from equation~(\ref{eq: Piecewise velocity})
compared with two other ideal vortex profiles \citep{smith1990numerical,grassi2018first}
(for the Grassi curve\citep{grassi2018first}, $\gamma=1.5$ was taken).
\textbf{b}, The vorticity calculated for the same profiles as \textbf{a}.
\textbf{c}, Vorticity gradient (in log scale), calculated for the
same three profiles. In addition, the minus of the $\beta$ profiles
are shown for the northern and southern poles of Saturn and Jupiter.
The 4 curves for $-\beta$ differentiate as the vorticity gradient
is normalized according to each polar cyclone, and as the length is
scaled by the radius of maximum velocity for the respective PC. The
points where the vorticity gradient curves cross the $-\beta$ curves
represent equilibrium. Here, $0$ in the $r/R$ axis represents the
pole. \label{fig:Vortex-velocity-profile}}
\end{figure}

\end{document}